\documentstyle[11pt,aaspp4]{article}  

\lefthead{Fusco-Femiano \& Menci}
\righthead {...}

\begin{document}

\newcommand{\lessim}{\ \raise -2.truept\hbox{\rlap{\hbox{$\sim$}}\raise5.truept
	\hbox{$<$}\ }}			

\title{Luminosity segregation from Merging \\
in Clusters of Galaxies}

\author{R. Fusco-Femiano}
\affil{Istituto di Astrofisica Spaziale, C.N.R.,\\
 C.P. 67, I-00044 Frascati, Italy}

\author{N. Menci}
\affil{Osservatorio Astronomico di Roma,
via Osservatorio, 00040 Monteporzio, Italy}

\begin{abstract}
We compute the evolution of the space-dependent mass distribution of galaxies 
 in clusters due to binary aggregations by solving a space-dependent 
Smoluchowski equation. From the solutions we derive the distribution of 
 intergalactic distance for different 
ranges of mass (and of corresponding magnitude). We compare the results 
 with the observed distributions, and find that the different degrees of 
luminosity segregation observed in clusters are well accounted for 
 by our merging model. In addition, the presence of luminosity segregation is 
 related to dynamical effects which also show up in different but 
connected observables, such 
  galaxy velocity profiles decreasing toward the center and X-ray measured   
$\beta$-parameters smaller than 1. 
We predict both luminosity segregation and the observables 
above (being a product of binary aggregations) to be inversely 
correlated with the core radius and with the galaxy velocity dispersion; we 
discuss how the whole set of predictions compares with up-to-date observations.

\end{abstract}


\section{ Introduction}

The dynamical evolution of galaxy clusters is currently believed to 
 go through two major phases: in the first, usually referred to as 
violent relaxation (Lynden-Bell 1967),  the evolution is controlled 
by a collective potential and results in a  Maxwell velocity distribution 
of galaxies;  in the second, the dynamics is dominated by two-body 
 processes, and binary (both elastic and inelastic) 
collisions drive the evolution. In fact, in this latter phase, for 
ordinary galaxy sizes and separations the 
 collision time scale is much less than the Hubble time. 

  Although a complete theoretical description of the two-body phase of 
 dynamical evolution of clusters is still lacking, observations, N-body 
 simulations and computation based on statistical 
methods (Monte Carlo and Fokker-Plank simulations) have concurred in 
 enlightening many dynamical properties of clusters in this stage. 
Such properties are characterized by a 
large cluster-to-cluster variance, and 
include the following: the presence of a {\it velocity bias}  
$b_v^2=\langle v^2 \rangle/\sigma^2 <1$ of the galaxy velocity dispersion 
$\langle v^2\rangle^{1/2} $ with respect to the dark matter's $\sigma$ 
(see N-body simulations by 
Carlberg \& Dubinski 1991; Evrard, Summers \& David 1994; Katz \& 
White 1993; Carlberg 1994; Summers, Davis $\&$ Evrard 1995); galaxy 
{\it velocity 
dispersion profiles decreasing toward the cluster center} (see observations 
 by Kent \& Sargent 1983; Sharples, Ellis, \& Gray 1988; 
Girardi et al. 1996);  {\it mass segregation}, i.e., the tendency of more 
massive galaxies to be located near the cluster center (see simulations by 
Roos \& Aarseth 1982; Farouki, Hoffman \& Spencer 1983)  
with the associated luminosity segregation (observed in several clusters, 
 see Rood et al. 1972; Oemler 1974; 
White 1977; Dressler 1978; Quintana 1979; 
 Sarazin 1980; Kent \& Gunn 1982; Oegerle, Hoessel \& Ernst 1986; 
 Binggeli, Tammann \& Sandage 1987; 
Dominguez Tenreiro \& Del Pozo-Sanz 1992; Stein 1996). 

Due to the large variance observed in the above effects,  
appreciable uncertainties exist about their dependence 
 on the characteristics of the system, although a general 
 trend in the sense of larger effects for smaller clusters might be 
inferred from the data,  when only relaxed cluster 
(with no prominent substructure or asphericity) are 
considered. 

A complete description of the post-virialization phase of galaxy cluster 
should be able to connect all the above effects and to explain the observed 
variance in terms of dynamical properties of clusters. 
In previous papers (Fusco-Femiano $\&$ Menci 1995, hereafter Paper I; 
 Menci \& Fusco-Femiano 1996, hereafter Paper II) we showed that 
the loss of kinetic energy in inelastic galaxy collisions  
(binary aggregations) in clusters with $\sigma\lessim 900$ km/s can 
 significantly change the velocity distribution of galaxies. The model 
 not only simultaneously accounts for the  
 velocity bias and for centrally decreasing velocity dispersion profiles, 
but also 
predicts a  correlation of such effects with the shape 
(the core radius) and the depth (the dark matter velocity dispersion) 
of the cluster potential wells in good agreement with observations. 
 In addition, the aggregation model can succesfully connect the above 
 dynamical effects to other observed properties of galaxy clusters (Paper I), 
such as 
the $\beta$-parameter (expressing the ratio of the galaxy orbital specific 
energy to the specific energy of the X-ray emitting 
plasma) and the Butcher-Oemler effect (see Cavaliere \& Menci 1993). 

Thus, aggregations seem to constitute a leading mechanism in the 
post-virialization phase of clusters  with velocity dispersion $\lessim  
 900$ km/s, while in larger clusters, they are 
highly suppressed due to the large galaxy relative velocities 
(see numerical results by Richstone \& Malmuth 1993). 
To further assess the role of galaxy merging in the two-body dynamical phase, 
 we address here the problem of luminosity segregation (hereafter LS). 
 This is expected to be generated when aggregations are effective, since 
 merging builds up larger galaxies mainly in the central regions, where 
the larger density favours binary aggregation. Thus, 
 we extend our treatement of galaxy inelastic collisions, 
based on the solution of a collisional  Boltzmann-Liuoville equation, to 
 include position-dependent mass spectra of interacting galaxies. 
  The prediction of our  model will be compared 
with observational results, focussing on the 
correlation of the segregation effects  
with the properties of the clusters such as 
the richness, the density distribution and the velocity dispersion. 
 Finally, we shall show how the merging model 
connects LS  with the dynamical 
 effects discussed above and with their X-ray counterparts. 

The paper is organized as follows. In sect. 2 we discuss the collisional 
Boltzmann equation and describe our solutions for the evolution of the 
 position-dependent mass 
distribution. In Sect. 3a we decribe the standard method 
we use for the  comparison with data, based on 
the auto-correlation function for galaxies of different mass. The 
comparison is performed in Sect. 3b.
Sect. 4 is devoted to discussion and conclusion.

\section { Radius-dependent mass distribution 
from binary aggregations} 

\subsection{The Boltzman Equation for 
Merging Galaxies}

The evolution with time $t$ of the distribution $f_t(M,{\bf r},{\bf v})$ 
of interacting 
galaxies with velocity ${\bf v}$ and mass $M$ at 
the position ${\bf r}$ inside 
the gravitational potential $\psi$ of a 
cluster can be described by the collisional Boltzmann equation. 
Assuming spherical symmetry, the latter can be written in 
spherical coordinates 
${\bf r}=(r,\theta,\phi)$ and ${\bf v}=(v_r,v_{\theta},v_{\phi})$ 
for the distribution $f(M,r,v_r,v_t)$ as follows 
(see, e.g.,  Saslaw 1985)
\begin{eqnarray}
\lefteqn{
 \partial_t f_t(M,r,v_r,v_t) +v_r\,{\partial f_t(M,r,v_r,v_t)
\over\partial r}+
\Big({v_t^2\over r}-{\partial\psi \over\partial r}\Big)-
2\,{v_r\,v_t^2\over r}\,{\partial f_t(M,r,v_r,v_t)\over 
\partial v_t^2} = } \nonumber \\
& & {1\over 2}\,
\int_o^M\,dM'\int_{-\infty}^{\infty}dv_r'\,\int_0^{\infty}\,dv_t'^2
 f_t(M',r,v_r',v_t')\,f_t(M-M',r,v_r'',v_t'')\,\Sigma(M',M-M',v_{rel})
\,v_{rel}+ \nonumber \\
& & 
-\int_0^{\infty}dM'\,\int_{-\infty}^{\infty}dv_r'\,\int_0^{\infty}\,dv_t'^2\,
 f_t(M,r,v_r,v_t)\,f_t(M',r,v_r',v_t')\,\Sigma(M',M',v_{rel})\,v_{rel}
\end{eqnarray}
where $v_t^2\equiv v_{\theta}^2+v_{\phi}^2$ is the square tangential component of velocity, and the gravitational cross section for interactions 
$\Sigma$ depends on the relative velocity ${\bf v}_{rel}\equiv 
{\bf v}'-{\bf v}''$. The velocity ${\bf v}''$ in the first 
integral in eq. (1) is related to ${\bf v}$ and to ${\bf v}'$ 
by the requirement of momentum conservation 
$M'\,{\bf v}'+(M-M')\,{\bf v}''=M\,{\bf v}$. 
Here we have assumed that the galaxies do 
not gain or loose mass via processes other than merging. 
To obtain a fully self-consistent description, eq. (1) should be 
complemented with the Poisson equation for the gravitational 
potential $\psi$ with a source term $\int dM\,dv_r\,dv^2_t\,
f_t(M,r,v_r,v_t)$, which 
includes the galaxy distribution itself. 
However, since we are interested in the post-virialization phase of 
cluster evolution where the potential is essentially fixed, 
 we shall assume a  King potential $\psi (r)$ 
and follow the evolution of the galaxy mass distribution 
at different radii. 

Since our aim is to probe the 
effectiveness of interaction in producing mass segregation, we 
indroduce some approximations (a discussion on them 
is given in the final section).
First, we assume the velocity distribution to be independent on 
the mass and on the spatial distribution of galaxies, so that 
 the distribution in eq. (1) can be factorized 
into a velocity distribution $p({\bf v})$ and a position-dependent 
mass distribution $N(M,r,t)$. 
Such approximation does not actually hold 
(see Paper I) but, as we discuss in the final
 Section, for our purpose in the present paper it is a 
{\it conservative} assumption. Second, we assume that the 
radial and tangential velocity distribution are mutually 
independent and both normally distributed. 
In this case, integration of eq. (1) over velocities leads to the 
following position-dependent Smoluchowski equation
\begin{eqnarray}
\partial_t N(M,r,t) & = & {1\over 2}\int dM'\,N(M',r,t)\,N(M-M',r,t) 
 \langle\Sigma (M',M-M')\,v_{rel}\rangle + \nonumber \\
& &  -\int dM'\,N(M,r,t)\,N(M',r,t)\langle\Sigma (M,M')\,v_{rel}\rangle
\end{eqnarray}
where the average $\langle\rangle$ is over the velocity 
distribution. 

The cross section is given by (Saslaw 1985): 
$\Sigma(M,M')=\epsilon(v_{rel}/v_g)\,
\pi\,(r^2+r'^2)\,
\Big[1+{v_g^2\over \,v_{rel}^2}\Big]$ 
where $r$ and $r'$ are the radii of the interacting galaxies 
(proportional to $M^{2/3}$) 
and $v_g\propto G(M+M')/R$ is the escape velocity at closest 
approach $R\approx (r+r')$. The efficiency $\epsilon$ is 
determined from N-body results (see Richstone \& Malmuth 1993) 
and is zero when $v_{rel}\gtrsim 3 v_g$, so that aggregations are highly 
suppressed in very rich clusters. 
It is convenient to express all 
quantities in terms of the {\it adimensional} 
mass $m\equiv M/M_*$ normalized to the 
characteristic mass $M_*$ (that corresponds to a galaxy with 
 characteristic luminosity $L_*$). From $r\sim(M/\rho)^{1/3}$, the
relation 
$r^2=r_{g*}^2\,m^{2/3}$ follows. Then, the cross section reads 
\begin{equation}
\Sigma(m,m')=\epsilon(v_{rel}/v_g)\,\pi\,r_{g*}^2
\,(m^{2/3}+m'^{2/3})\,
\Big[1+(m^{2/3}+m'^{2/3})\,v^2_{g*}/v^2_{rel}\Big]
\end{equation}
where $r_{g*}$ and $v_{g*}$ are the radius and the 3-D internal velocity 
dispersion of a $L_*$ galaxy, respectively.

\subsection{ Initial Conditions}

We assume the galaxy distribution to be {\it initially} 
(after the cluster formation and virialization) 
factorized in a mass distribution $P(m)$
times a King spatial profile, which will be subsequently 
mixed up by the 2-body dynamical evolution. 
Then 
\begin{equation}
N(M,r)_{t=0}={n_o\over (1+x^2)^{3/2}}\,P(m)~,
\end{equation}
where we take for $P(m)$ the Press \& Shechter shape 
$P(m)=m^{a-2}\,e^{-b^2\,\delta_c^2\,m^{2a}/2}$ (the index 
$a$ depends on the spectrum of cosmological 
perturbations and is in the range 0 - 0.3 at the  scale
of galaxy clusters) and $x=r/r_c$ is the distance from 
the cluster center in units of the core radius $r_c$ 
of the King profile. The constant $n_o$ is taken as to yield the
total number $N_{tot}$ of galaxies inside the cluster virial 
radius $R_v$ (from the virial theorem $R_v=G\,M/3\sigma$); thus 
\begin{equation}
n_o={N_{tot}\over 4\,\pi\,r_c^3\,I_R\,I_M}~,
\end{equation}
where $I_R\equiv \int_0^{R_v/r_c}dx\,x^2/(1+x^2)^{3/2}$ and $I_M\equiv
 \int_0^{\infty}\,dm\,P(m)$ are the adimensional 
integrals of the initial 
spatial and mass distributions, respectively.

\subsection{Numerical Solutions}

To integrate eq. (2) we first write it in a completely 
adimensional  form for the normalized $r$-dependent mass 
distribution $n_t(m,r)\equiv N(m,r,t)/n_o$.   We define the 
 adimensional time variable 
$\tau\equiv t/t_{cr}\approx 2\,10^9\,{\rm yr}\,
\big[R_v/1\,{\rm Mpc}\big]\,\big[\sigma/10^3\,{\rm km/s}\big]^{-1}$ 
in terms of the 
the cluster crossing time $t_{cr}\equiv 2\,R_v/\sigma$.  
The adimensional 
velocities $\tilde{v}\equiv v/\sigma$ are normalized to the dark matter 
velocity dispersion $\sigma$. The corresponding adimensional  
interaction rate $\eta (m,m')=n_o\,\Sigma\,R_v\,\tilde v$ 
can be computed from eq. (3) and (4). Then, the Smoluchowski eq. 
for the normalized mass distribution $n_{\tau}(m,r)\equiv N(M,r,\tau)/n_o$ 
can be recast in the form 
\begin{mathletters}
\begin{eqnarray}
\partial_{\tau} n_{\tau}(m,r) & = & 
 {1\over 2}\,\int_0^m\,dm'\,n_{\tau}(m',r)\,n_{\tau}(m-m',r)\,
\langle\eta(m',m-m')\rangle+ \nonumber \\
& &
-\int_o^m\,dm'n_{\tau}(m,r)\,n_{\tau}(m',r)\,\langle\eta(m,m')\rangle \\
\eta(m,m') &= &{1\over 2}\,
{R_v\over r_c}\,{r_{g*}^2\over r_c^2}
\,{N_{tot}\over I_R\,I_M}\,\tilde v_{rel}
\,(m^{2/3}+m'^{2/3})\,
\Big[1+(m^{2/3}+m'^{2/3})\,\tilde v^2_{g*}/\tilde v^2_{rel}\Big]
.
\end{eqnarray}
\end{mathletters}
From the form (6) it is evident how 
 (for constant $R_v/r_c$ ratio) 
 the effect of aggregation is larger for clusters with small 
core radius (galaxies in the center are more concentrated) 
and with a larger number of galaxies $N_{tot}$. 

The average of the aggregation rate in eq. (6b) 
\begin{equation}
\langle \eta\rangle\equiv \int\,d\alpha
\int_o^{ {|\bf\tilde{v}_1}-{\bf\tilde{v}_2}| =3\,\tilde{v}_g}
d\tilde{v}_1\,\tilde{v}_1^2\,p(\tilde{v}_1)\,d\tilde{v}_2\,
\tilde{v}_2^2\,p(\tilde{v}_2)\,
\eta({\bf\tilde{v}_1}-{\bf\tilde{v}_2}|/\tilde{v}_g)~,
\end{equation}
is over the velocities $\tilde{v}_1$ and $\tilde{v}_2$ 
 (normalized to the dark matter velocity dispersion $\sigma$) 
of galaxies colliding with relative angle $\alpha$; the 
   condition $|{\bf\tilde{v}_1}-{\bf\tilde{v}_2}| \leq 3\,\tilde{v_g}$
 accounts for the efficiency $\epsilon (v_{rel}/v_g)$. 
We assume the distribution of velocities 
$p(\tilde{v})=(1/2\pi)^{-3/2}\,e^{-\tilde{v}^2/2}$ to be Gaussian, 
as expected after violent relaxation (Lynden-Bell 1967). 
Note that, for clusters with $\sigma\lesssim 900$ km/s, eq. (7) yields 
 significant averaged aggregation rates $\langle \eta\rangle$ 
 assuming a 3-D internal velocity dispersion $v_{g*}= 300$ km/s for an 
 $L_*$ galaxy with $r_{g*}=60\,h^{-1}$ kpc
\footnote{In the text we adopt $h=0.5$ for the Hubble constant 
$H_o=100\,h$ km/s/Mpc.  }.
The adopted value of $v_{g*}$
 correspond to a circular velocity of $\approx 220$ km/s;  
 such value is consistent with that derived from the Faber-Jackson relation 
 for an $L_*$ galaxy,  
and with the measurements by Tonry \& Davis (1981); Dressler (1984); 
 Dressler (1987). The adopted value of $r_{g*}$ 
 (which refers to the dark halo of  an $L_*$ galaxy) 
 is a conservative one, when compared with observational results from 
absorption lines, measured by Steidel (1995); Lanzetta et al. (1995); 
Barcons, Lanzetta \& Webb (1995). 

Equation (6) is integrated up to $\tau=5$ with time increments 
$\Delta \tau=1/500$ and mass step $\Delta m=1/500$, from a 
 minimum mass $m_{inf}=10^{-2}$ to a maximum mass $m_{sup}=10^2$ 
(integrating up to larger times do not affect sensitively our results). 
When aggregations are effective, 
the final mass distribution will be changed from the initial 
one only in the central core where the galaxy density is larger 
and binary aggregations are favoured. Thus, in the core larger 
galaxies will form via binary merging, while the initial 
mass distribution remains unchanged in the outer regions. 
The evolution of the mass distribution at different radii in a 
 typical cluster (with $N_{tot}=1000$, $\sigma=800$ km/s and 
$r_c=250\,h^{-1}$ kpc) is shown in figure 1. The distribution 
flattens in the central region due to the disappearence of 
small galaxies which aggregate to form larger ones. Since aggregations 
 between galaxies cause a loss of orbital kinetic energy, we expect such 
effect to be correlated with galaxy velocity dispersions smaller 
 in the central regions, i.e.,  with velocity profiles falling toward the 
 center (as 
we discussed in Paper II); this is actually the case, as is shown in figure 2.
A further effect is that brighter galaxies (which form from 
 mainly in the central, denser regions) will 
have smaller relative separations. 
 This latter effect is that observed in several clusters, 
as we discuss in the next session. 

The strenght of the above effects depends on the cluster properties, which, 
in our model, enter only through $N_{tot}$, $\sigma$ and $r_c$ as is 
shown by eqs. (6).  
E.g., for given $N_{tot}$ and $\sigma$, in 
 clusters with large $r_c$ merging will be less effective (see the 
merging rate in eq. 6b) 
because the total number of galaxies is spread out in a larger region.

\section{Comparison with Observations}

\subsection{Method}

In the literature (Capelato et al. 1980; Dominguez-Tenreiro \& 
del Pozo-Sanz 1988) 
the LS  has been quantified in terms of the 
cross correlation function 
\begin{equation}
\Pi_{a}(s)=\int_V\,d\phi\,d^2{\bf r}\,n_a({\bf r})\,
n_a({\bf r}+{\bf s})
\end{equation}
between densities of galaxies  
separated by a distance ${\bf s}$ in a given magnitude range $[a]$
($\phi$ is the angle between ${\bf r}$ and ${\bf s}$), in a region $V$. 
The distance distribution function for  pairs in a given class 
is then given by 
\begin{equation}
P_{a}(s)ds=2\,\pi\,sds\,\Pi_{a}(s)~.
\end{equation}
If  the position of the peak in the distribution $P(s)$ changes 
 depending on the magnitude class $[a]$
a LS  is present. The average separation 
 of galaxies in the magnitude class $[a]$ derived from eq. (8) reads 
\begin{equation}
\lambda_a=\int_0^{S_{max}}\,ds\,s\,P_a(s)/\int_0^{S_{max}}\,ds\,P_a(s)~,
\end{equation}
where $S_{max}$ is the maximum intergalactic distance. Galaxies belonging 
 to a class $[a]$ will be called ``segregated'' with respect 
to those in the class $[a']$ if $\lambda_a<\lambda_{a'}$. 

Such an effect has been measured in several clusters (Capelato et al. 1980; 
 Dominguez-Tenreiro \& del Pozo-Sanz 1988; 
Yepes, Dominguez-Tenreiro \& del Pozo-Sanz 1991; 
Yepes \& Dominguez-Tenreiro 1992)
and a fit to the observed $P(s)$ for different classes of 
magnitudes has been given by the same authors. 
We consider all the clusters where such an analysis has been 
performed, except for Perseus, whose prominent substructure (Gallagher, Han 
$\&$ Wyse 1996)) makes it 
 too complex to be describable in the framework of  
 our model. They are reported in Table 1 together with their 
core radius, velocity dispersion and observed number of galaxies 
inside a distance $R_{max}$. In the same table are listed the 
 magnitude ranges $[a]$ for which we compute the correlation 
function. 

To compare with such observational material 
we proceed as follows: 
\newline

$\bullet$ For each observed cluster, we compute the number of galaxies 
$N_{tot}$ enclosed inside $R_V$ 
for the whole mass range $0.01<M/M_*<10$ (used in the 
numerical integration of eq. 6), corresponding to 
 a luminosity range $0.01<(L/L_*)^{\gamma}<10$ for a $M/L\propto 
 L^{\gamma -1}$ (here we take $\gamma=1$, but see discussion in Sect. 4 for 
 the effect of changing $M/L$).

In practice, $N_{tot}$ 
 is computed extrapolating the observed number $N_{obs}$ of galaxies (inside 
 a radius $R_{max}$, see Table 1)  
both in space (up to $R_v$, using a King profile) and in luminosity 
(for the whole luminosity range discussed above, 
using a Shechter luminosity function). 

The resulting $N_{tot}$ is given as an input for the solution of eq. (6). 
together with the cluster core radius $r_c$ and the dark matter velocity 
 dispersion $\sigma$. The latter is derived from observed galaxy velocity 
dispersions (assuming no velocity bias) or, when the latter are not available, 
from the X-ray temperature $T=(\mu m_H/k\,\beta)\sigma^2$ 
(when measures of $\beta$ are not available, we shall assume $\beta=1$).
The resulting values (with the 
references to the corresponding observations) are given in Table 1.
 The reported $\sigma$ are affected by uncertainties $\Delta\sigma/\sigma< 
20 \%$ due to intrinsic errors in the measurements of velocity dispersions 
or X-ray temperatures and 
(when the estimate of 
$\sigma$ is obtained from $T$ with no available measurements of $\beta$) 
 to the indetermination of $\beta$. However, we stress that 
 errors in $\sigma$ (as well as those on $N_{obs}$) 
do not affect sensitively the LS effects 
resulting from our model. A quantitative discussion of the 
 effect of variations of all 
the input parameters is given in Sect. 3.3.

$\bullet$ For each cluster the $r$-dependent mass 
distribution is found integrating numerically eq. (6). 

$\bullet$ We divide the computed 
mass distribution at each radius according to the classes 
of apparent magnitudes (see Table 1) which have been used in the 
analysis by 
Capelato et al. 1980;  Dominguez-Tenreiro \& del Pozo-Sanz 1988; 
Yepes et al. 1991; Yepes \& Dominguez-Tenreiro 1992. 
To pass from magnitude to mass ranges we use the $M/L$ ratio 
discussed above. 

$\bullet$ We compute the distance distribution $P(s)$ 
resulting from our model and compare it with the fit to 
observational results found in the literature. 
For each cluster,  
 the average separation $\lambda_a$ corresponding to $P_a(s)$ is 
computed for all the magnitude classes $[a]$ 
and compared with the observed values. 

When the cluster characteristics are such as to make aggregations effective, 
 larger galaxies form preferentially in the central, denser regions 
(see Sect. 2 and fig. 1) where the intergalactic separation are smaller. 
 In this case, the distributions $P_a(s)$ will be peaked at smaller 
 separations for brighter magnitude ranges $[a]$. Such shift of the 
 peak with $[a]$ can be expressed by the ratio
\begin{equation}
b_{\lambda}\equiv \lambda_1/\lambda_3~
\end{equation}
of the average distances of the brightest class to that of the faintest class. 

\subsection{Results}

The distributions $P_a(s)$ for the different clusters are shown in fig. 3. 
The ratio $b_{\lambda}$ 
 is shown in Table 2 for our predictions and for the corresponding 
observations. The agreement with observations is remarkable.  
 The different degrees of segregation (expressed by values $b_{\lambda}<1$) 
 observed in the sample is well accounted for by our merging model, and 
 is directly related to the cluster characteristics as follows: 
for a given total number of galaxies $N_{tot}$, 
clusters with small core radii have denser central regions, so 
that aggregations are more effective and segregation is enhanced. 
The lack of LS in A2111 is explained in terms 
of large core radius coupled with a limited total number of 
galaxies. As a confirm to such an interpretation, we observe that 
 the more pronounced segregation takes place in A2670 which is 
characterized by the smallest core radius in the sample. However, LS can occur also
in clusters with large $r_c$ if the total number of galaxies $N_{tot}$ is large
enough or if $\sigma$ is very low. In fact, the cluster 0004.8-3450 shows a 
significant LS due to $N_{tot}\simeq 2000$, while the segregation in the Fornax
cluster is mostly due to its very low velocity dispersion $\sigma$=320 km/s. 

Note the peculiarity of cluster A2218. The model is in very good agreement 
for what concerns the observed LS  of the two 
brightest magnitude classes with respect to the third one. 
 However, the real data show that the two brightest 
 classes are characterized 
by an anti-segregation between them, 
which is not accounted for by our model. We attribute such mismatch to 
  substructures/anisotropies which 
our model (based on the schematic assumption of isotropy) cannot reproduce. 
In fact, recent analysis (Squires et al. 1996) of A2218 
indicates the presence of collision of subclumps, with 
associated elongated structures in the plasma disposition. 

 As observed above, our model predicts  the aggregating galaxies 
 to loose part of their kinetic energy. The brightest galaxies in a 
cluster showing segregation are then expected to have 
 velocity dispersion profiles decreasing toward the center, where the larger 
 density favours aggregations. 
The computed result for 
 the galaxies belonging to the magnitude classes 1 and 2 in 
A2670 (see figure 4) confirm such expectation and are consistent 
with the available data for such cluster (Sharples, Ellis, \& Gray 1988; 
 see also Yepes \& Dominguez-Tenreiro 1992). The same calculation for A2111 
 (see figure 4) shows no positive gradient in the profiles,  
 which is the counterpart of the lack 
of luminosity segregation. Actually, both effects are tightly 
connected in our model.

\subsection{Varying the Input Parameters}

Here we discuss the effects of variations of the input parameters 
with respect to the reference values in Table 1. 
 The segregation parameter $b_{\lambda}$ decreases (indicating larger 
segregation) for increasing $N_{obs}$, and for decreasing 
$\sigma$ and $r_c$ (i.e., for increasing merging efficiency). 
 However the variations with $N_{obs}$ and $\sigma$ are very mild. This 
makes our results {\it robust} with respect to the errors associated to those 
parameters: a 20 $\%$ error in $N_{obs}$ or in $\sigma$ 
 results in $\Delta b_{\lambda}/b_{\lambda}< 3 \% $.  
The errors in $r_c$ are more 
important: $\Delta r_c/r_c=20\%$ yields $\Delta b_{\lambda}/b_{\lambda}
 < 12 \% $. 

The results for LS do depend on the $M/L$ ratio which 
for the sake of simplicity, we assumed to be constant.  
However, the main results presented here holds also for 
$M/L\propto L^{\gamma -1}$ with $3/4\leq \gamma\leq 4/3$. 
This is illustrated in 
figure 5, where we show the distance distribution functions (for 
 the parameters of cluster A2670) derived from the {\it same} dynamics 
 (i.e., with the same mass segregation) but with $\gamma=3/4$ and 
$\gamma=4/3$.

\section{Conclusions}

We have shown that a detailed model for the dynamics of  galaxies 
 aggregating in the potential wells of clusters predicts luminosity 
segregation (LS) effects of the kind observed in real clusters. 
 The correlation of the strenght of the effect with the properties of the 
clusters predicted in our model is in agreement  (see figure 3 and Table 2)
 with that observed for the   
limited sample of clusters for which LS has been subject to accurate 
quantitative measurements. 
 In particular, we predict the effectiveness of 
aggregations, and hence the degree of LS, to be 
 directly correlated with the number of galaxies in the cluster and inversely 
correlated with the core radius and with the velocity dispersion (see eq. 6b).

 The results do not depend on the detail of the initial mass distribution of 
 galaxies in clusters, which we assume to have a Press \& Shechter form 
with spectral parameter $a=-2$; such independency is due
  to the properties of the asymptotic solution 
of the Smoluchowski equation (describing 
 the evolution of the position-dependent galaxy mass function in our model) 
and can be traced back to the non-linear nature of such equation. 

Our results are robust with respect to uncertainties in the input 
quantities and to the adopted $L(M)$ , as shown in Sect. 3.3.
As for our assumption of fixed galaxy velocity distribution,  
 this does not hold when aggregations are effective (see Paper I and II). 
 However, the merging-induced shift of $\sim (10-15)\%$ of the velocity 
dispersion toward smaller values  {\it increases} the efficiency of 
aggregations, so that our assumption is actually conservative. 
We have re-run our computation for shifted velocity distributions and found 
 results almost indistinguishable from those presented here. 
Finally, we 
stress that no attempt of parameter optimization has been performed. 
An even better agreement could be found if the cluster parameters were 
 suitable tuned. 

As for the big picture of the evolution of cluster in the 
two-body dynamical phase, 
our model focus on  the effects of inelastic collisions not  
 considered in previous works on this subject. In particular, the 
Fokker-Plank approach 
 by  Yepes \& Dominguez-Tenreiro (1992) considers only elastic collisions 
by means of a  ``mean field'' approximation with the input parameters 
 chosen  from a grid of models to show that, 
within the set of models, it is possible to 
match the observed segregation effects. 

Here we solve the collisional Boltzmann equation including inelastic 
collisions, 
 using, for the input model parameters, the {\it measured} values. 
Though the latter are subject to errors, we showed (in Sect. 3.3) that 
 the model is robust to uncertainties $< 12 \%$ in the parameters. 
 Our results show that inelatic collisions produce appreciable 
dynamical effects 
 for clusters with one-dimensional velocity dispersion $\lessim 900$ km/s. 
 Such effects show up in different but connected observables: 
 the velocity bias (due to the {\it average} loss of kinetic energy 
in inelastic collisions) $b_v\approx 0.8-0.9$ 
 can be observed in X-rays in the form of $\beta$-parameter (see Cavaliere \& 
 Fusco-Femiano 1976)  $\beta=b_v^2<1$; centrally rising velocity 
 dispersion profiles (due to the {\it differential} loss of kinetic energy at 
different radii) are now being measured with great accuracy in the optical 
 (Girardi et al. 1996); 
different average separations of massive galaxies with 
respect to the faint ones $b_{\lambda}\approx 0.8-0.9$ (see eq. 5), 
i.e., luminosity segregation (due to the 
  differential mass growth from aggregations at different radii)  have been 
 measured in different clusters (see references cited in this paper).

We note that, when interpreted in terms of merging-driven evolution, 
 {\it all} the above effects are predicted to have the {\it same} 
dependence on the cluster parameters, i.e., to be larger for clusters 
with smaller core radii $r_c$ and galaxy velocity dispersions $\sigma$, 
although the strength of the $\sigma$-dependence is mild for 
LS effects. 

The observational tests for such predictions are critically affected by the 
presence of clusters with anisotropies and/or substructures in the 
 observational sample. 
An inverse correlations of the $\beta<1$-effect with $r_c$ has been found   
by, e.g., Jones \& Forman 1984, while the anti-correlation with $\sigma$ 
 has been pointed out by 
 Kriss et al. (1983); Jones \& Forman (1984); Edge \& Stewart (1991); 
  Bird, Mushotzky \& Metzler (1995); Jones et al. (1997) 
 but has not been confirmed by the analysis by Lubin \& Bahcall (1993) and by 
Girardi et al. (1996). 

 For velocity dispersion profiles decreasing toward the center, the 
 observational situation is still unclear. An anti-correlation with $\sigma$ 
 has been inferred (see Paper II)
from the analysis by Girardi et al. (1996) of a 
sample of 37 clusters, when clusters with prominent substructures are excluded;   
 however, the detection of such correlation from the data 
(see also den Hartog, \& Katgert, 1996) is made difficult 
 by the presence of anisotropies and/or substructures, which can hurt or 
destroy the effect of the inelastic collision 
in the two-body relaxation phase. 

As for the LS, 
the mild (inverse) 
dependence on $\sigma$ of the LS effect from merging makes difficult 
to observe such correlation. However, in our model,  the strong inverse 
correlation of LS with $r_c$ predicted by our model is confirmed 
the data analysis 
by Yepes, Dominguez-Tenreiro \& Del Pozo-Sanz (1991)
 on the very limited sample of clusters. 
LS data for a larger sample of cluster with measured $r_c$ would definitely 
clarify the issue.

Finally, we note that the  correlations between 
 different but connected observables predicted by the aggregation model makes 
it 
 testable already at the present stage of observational capabilities. 
 Further observational progress (in particular in measuring in detail 
velocity dispersion profiles and X-ray temperatures) can definitely 
 probe the predictions of the merging picture, thus 
 assessing  the role of aggregations in the  dynamical evolution of clusters. 

\acknowledgments 
We thank the referee for keen suggestions and comments.

\newpage

\figcaption[fig1.eps]{The position dependent mass distribution obtained from 
the solution of eqs. 6, for a cluster with the parameters given in the text (arbitrary
units are used for $r$ and $M$).
 Top panel refers to the initial condition, while the 
 bottom panel to the evolved distribution at $t=5\,t_{cr}$. Note the 
flattening at the  low-mass end in the central region}

\figcaption[fig2.eps]{The corresponding integrated velocity dispersion 
 calculated as shown in Paper II. Note the decrease toward the central region, 
 corresponding to the loss of orbital energy by aggregations occurring in the  
 cluster center}

\figcaption[fig3.eps]{The computed intergalactic distance distribution 
 for the clusters with parameters listed in Table 2. The solid curve refers to 
 the brighter magnitude class, while the dotted line to the fainter. 
All the curves have been computed for a constant $M/L$} 

\figcaption[fig4.eps]{The integrated velocity dispersion profiles computed 
 as shown in Paper II, for the cluster A2670 and 
 A2111} 

\figcaption[fig5.eps]{The computed intergalactic distance distribution 
 for the cluster A2670, for different values of the $M/L\propto L^{\gamma -1}$ 
 (see text) ratio. 
Top panel refers to $\gamma=3/4$, while bottom panel to $\gamma=4/3$.}
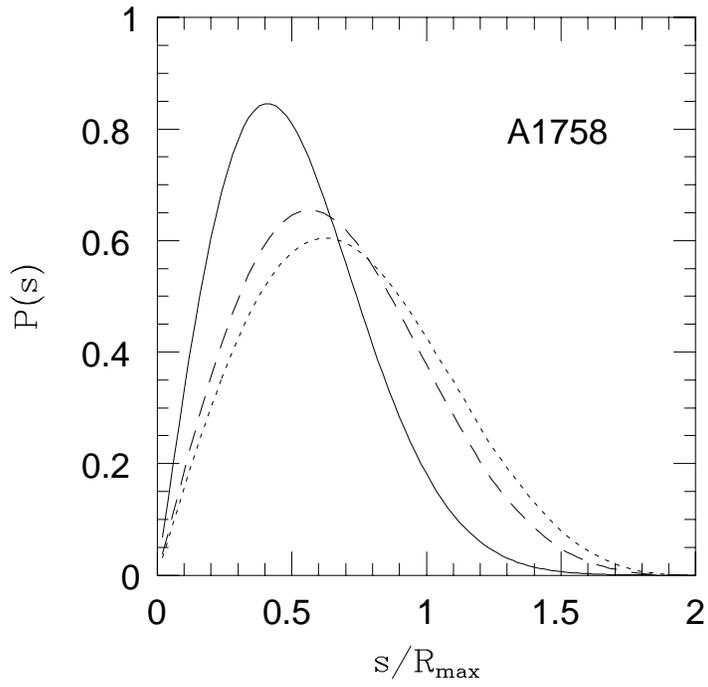
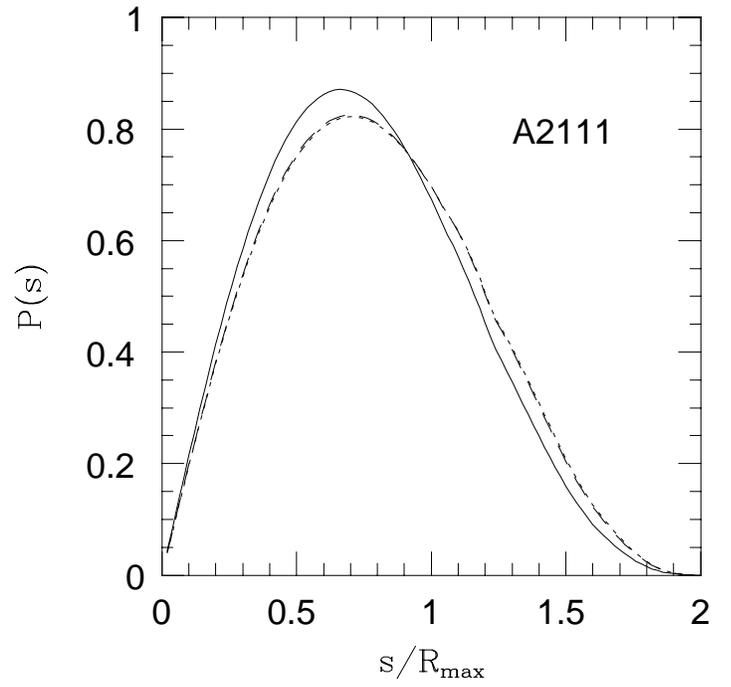
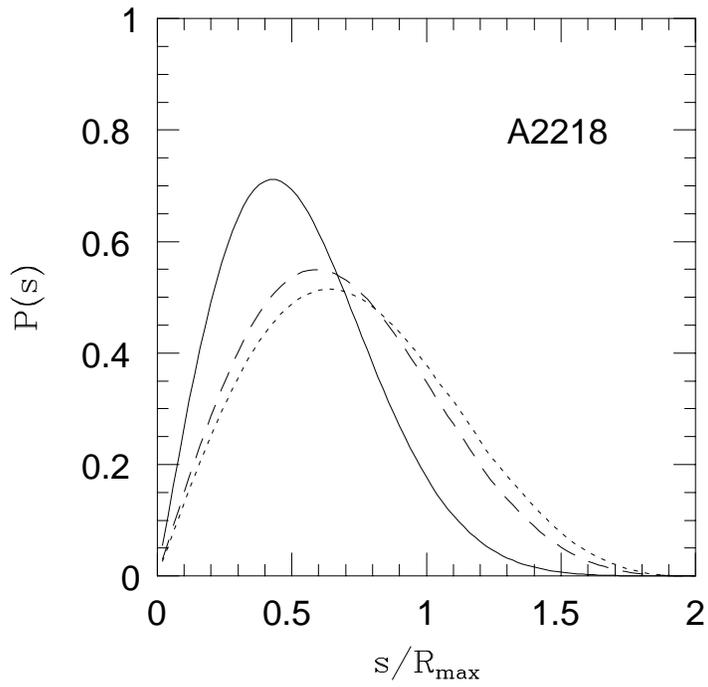
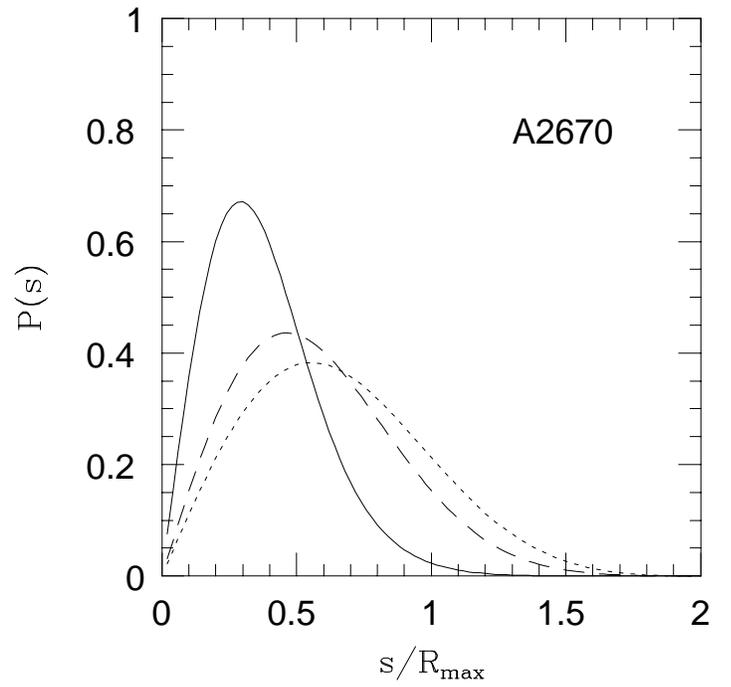
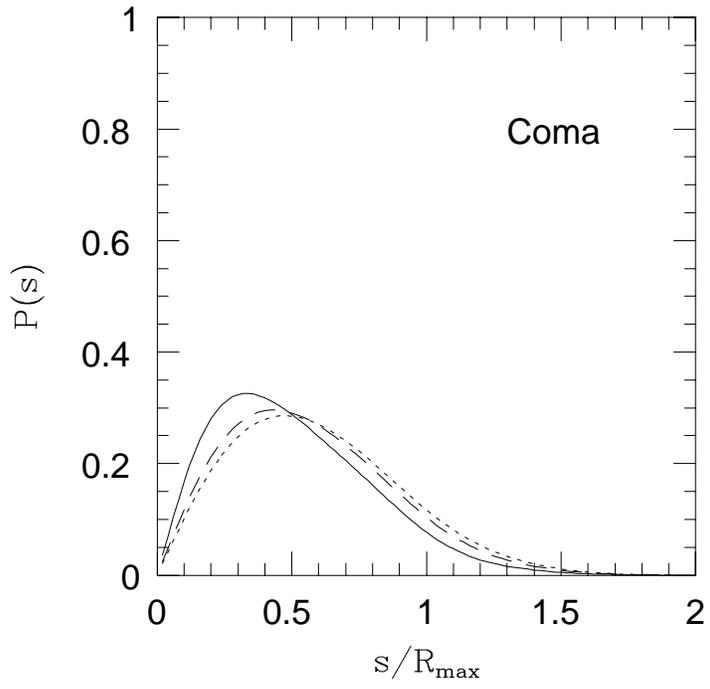
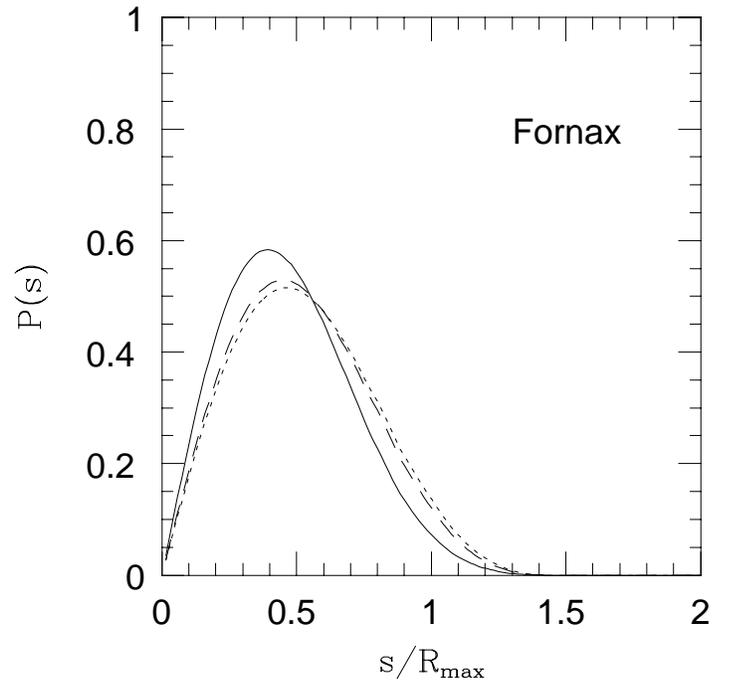
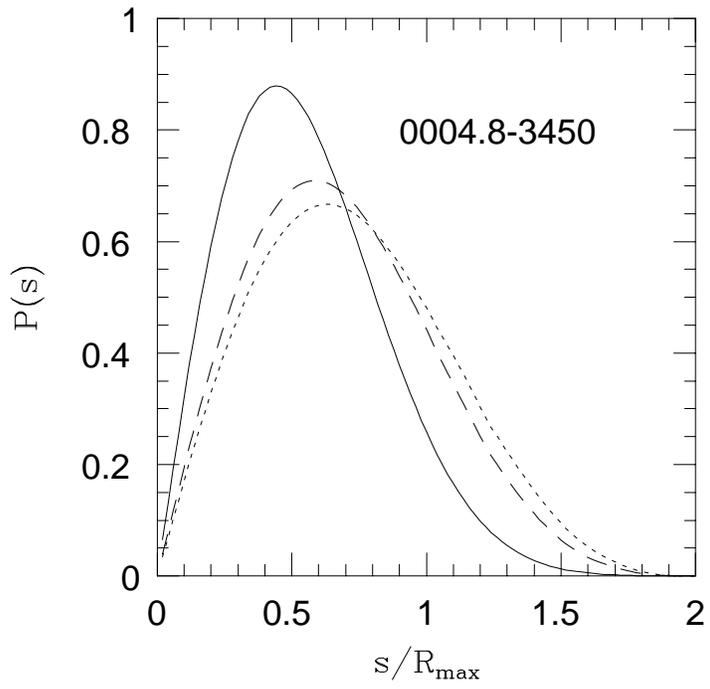
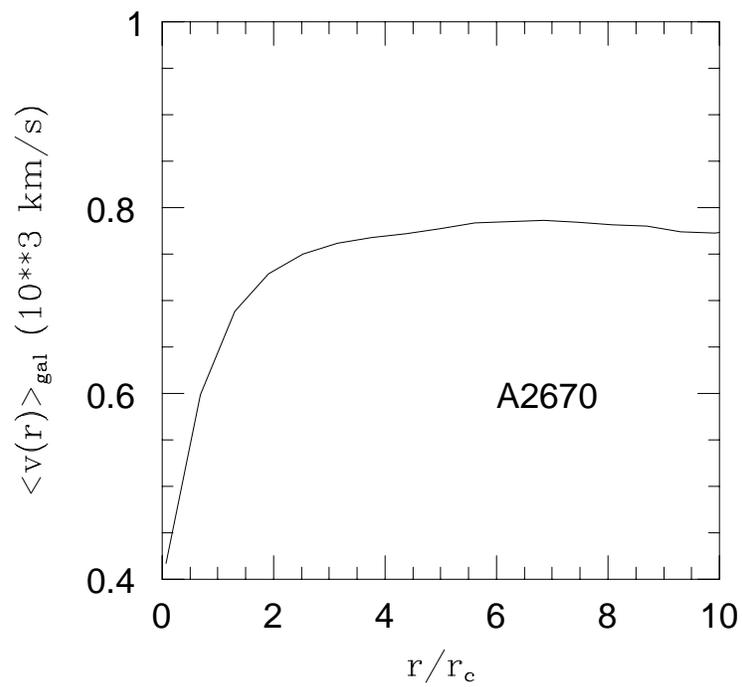
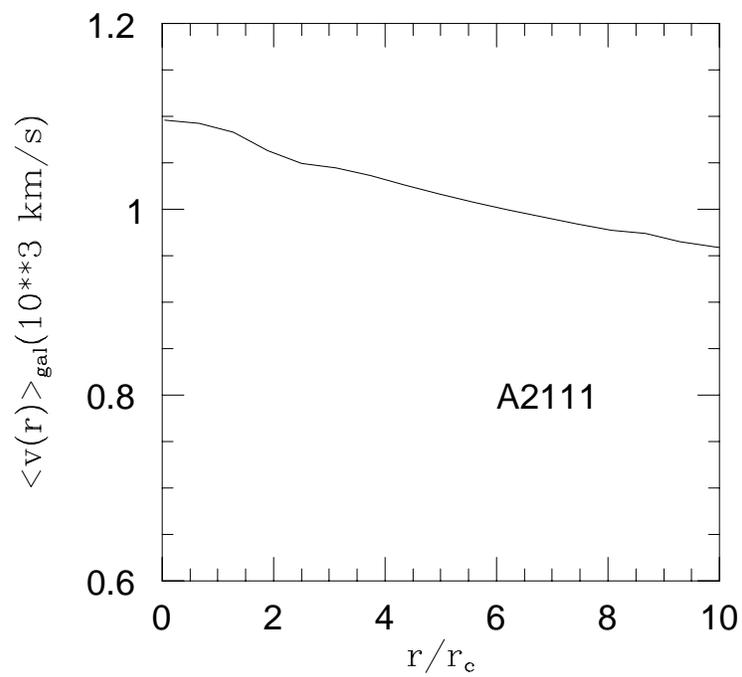
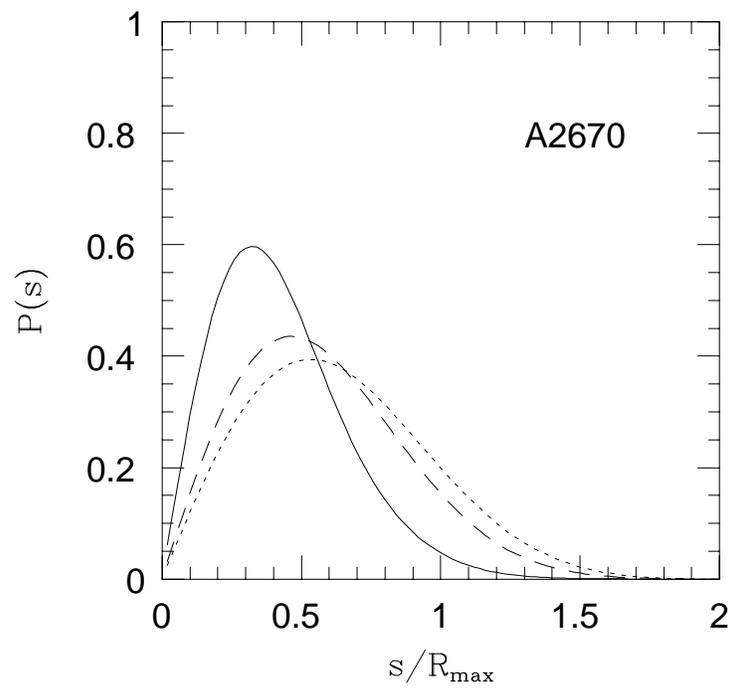
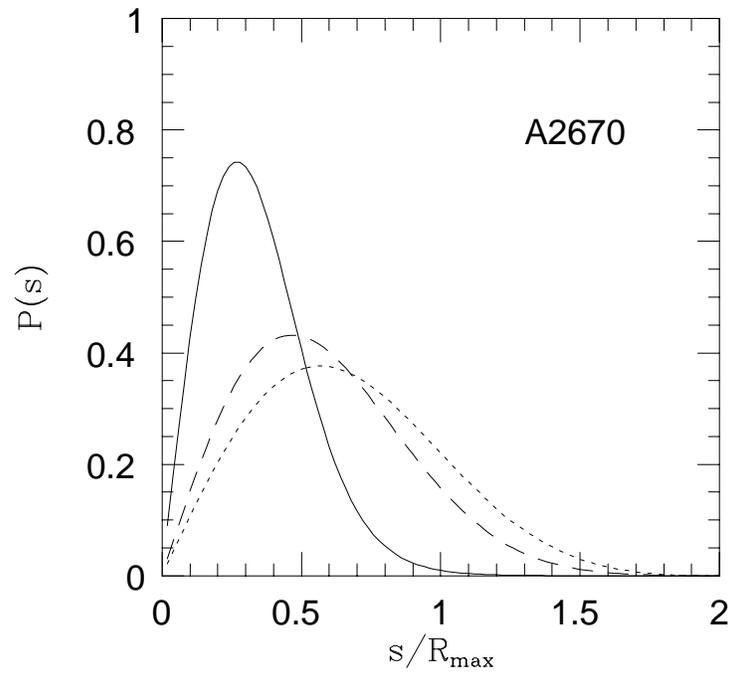
\end{document}